\documentclass[preprint,onecolumn,nofootinbib]{revtex4}
\usepackage{graphicx}
\pdfoutput=1
\usepackage[colorlinks=true,linkcolor=blue,
urlcolor=blue,filecolor=black,citecolor=red,
pdfstartview=FitV,pdftitle={},pdfsubject={},
pdfkeywords={},pdfpagemode=None,bookmarksopen=true]{hyperref}
\usepackage{epsfig}
\usepackage{amsmath}
\usepackage{amsfonts}
\usepackage{amssymb}
\usepackage{color}%
\usepackage{dcolumn}
\providecommand{\U}[1]{\protect\rule{.1in}{.1in}}

\newcommand{\f}{\begin{equation}}
\newcommand{\ff}{\end{equation}}
\newcommand{\fa}{\begin{eqnarray}}
\newcommand{\ffa}{\end{eqnarray}}

\begin{document}
\title{A Novel Insulator by Holographic Q-lattices}
\author{Yi Ling $^{1,3}$}
\email{lingy@ihep.ac.cn}
\author{Peng Liu $^{1}$}
\email{liup51@ihep.ac.cn}
\author{Jian-Pin Wu $^{2,3}$}
\email{jianpinwu@mail.bnu.edu.cn}
\affiliation{$^1$ Institute of High Energy Physics, Chinese Academy of Sciences, Beijing 100049, China\ \\
$^2$Institute of Gravitation and Cosmology, Department of Physics,
School of Mathematics and Physics, Bohai University, Jinzhou 121013, China\ \\
$^3$ State Key Laboratory of Theoretical Physics, Institute of Theoretical Physics, Chinese Academy of Sciences, Beijing 100190, China
}
\begin{abstract}
We construct a bulk geometry with Q-lattice structure, which
is implemented by two gauge fields and a coupling between the
lattice and the Maxwell field. This gravity dual model can
describe a novel insulator which exhibits some key features
analogous to Mott insulator. In particular, a hard gap in
insulating phase as well as vanishing DC conductivity can be
simultaneously achieved. In addition, we discuss the
non-Drude behavior of the optical conductivity in low frequency
region in insulating phase, which exhibits some novel
characteristics different from ordinary Mott insulator.
\end{abstract}
\maketitle

\section{Introduction}

Metal-insulator transition, which exhibits very appealing and
peculiar properties in the electronic conductivity, is a
fundamental issue in condensed matter physics
\cite{Mott:1974,Imada:1998,Dobrosavljevic:2012,Basov:2011}. In
particular, Mott insulator is a typical many-body system in
which the transport of electrons is inhibited not due to the
full filled band but rather to strong local electronic
correlations \cite{Mott:1930s}. For such strongly correlated
system, the conventional methods are ineffective and the basic
principle behind them is fuzzy. AdS/CFT correspondence
provides a powerful tool and new paradigm for understanding
strongly correlated systems
\cite{Maldacena:1997,Gubser:1998,Witten:1998}. In holographic
approach various novel localization mechanisms have been
implemented with the lattice deformation in bulk geometry, which
is usually solvable such that it provides an intuitive geometric
scenario from gravity side for the dual system with strongly
correlated electrons. When the operator breaking the translational
symmetry becomes relevant in the IR region, a
metal-insulator transition will happen in the dual field
theory \cite{DonosHartnoll,Donos:2014oha,Donos:2013eha,Donos:2014uba,GouterauxHCA,Ling:2014saa,BaggioliROA}.
Different lattice structure results in different localization
mechanism such that we have a diverse landscape of insulating
phase in a holographic scenario, for instance, Peirels insulator
\cite{Ling:2014saa}, polaron-localization insulating phase
\cite{BaggioliROA} or other novel insulating phases exhibiting
peculiar characteristics
\cite{DonosHartnoll,Donos:2014oha,Donos:2013eha,Donos:2014uba,GouterauxHCA},
some of which are similar to those found in condensed matter.

In this paper, we will construct a holographic model which is
dual to a novel insulator that shares some important characteristics of Mott insulator.
In this route the pioneer work on the formation of hard gap has
appeared in \cite{Kiritsis:2015oxa,LingEPA}. In
\cite{Kiritsis:2015oxa}, the momentum relaxation is introduced by
axion field $\chi_i=k x_i$ in some gapped geometries
\cite{CharmousisZZ,GouterauxCE,GouterauxYR} such that a hard gap
in optical conductivity and an insulating ground state can be
obtained. But the thought experiment originally proposed by Mott
is not transparent in this scenario due to the absence of periodic
lattice structure. In Mott thought experiment, when the repulsive
force between two electrons is taken into account, a metal could
transit into an insulator with the increase of the lattice
constant, simply because the hoping ability of electrons decays.
In our recent work \cite{LingEPA}, we introduce a coupling between
the lattice and the electromagnetic field over a Q-lattice
background which exhibits a manifest periodic structure
\cite{Donos:2013eha}. We find our model can not only implement the
Mott thought experiment explicitly, but also exhibit a hard gap
evidently in optical conductivity when the coupling parameter is
relatively large. However, in this framework we find the dual
system always stays in a novel metallic phase in zero temperature
limit whenever a hard gap is formed. We conjecture it is a doped
system where umklapp scattering is frozen at zero temperature. As
a result, a ground state corresponding to a Mott insulating phase
at zero temperature is not achieved in \cite{LingEPA}. Based on
our previous work, in this paper we intend to report a substantial
progress to \cite{LingEPA}. We will introduce a gravity dual model with two $U(1)$
gauge fields. This two-gauge formalism has been introduced into holographic
models for a long time, with a long list of publications, for
instance, in \cite{Donos:2012yi,Donos:2013gda,Ling:2014saa,DonosHartnoll}. It originally comes from the
top-down construction of holography based on the low energy limit
of superstring or M theory \cite{Cvetic:1999xp,Donos:2012yi}.
In our current work, the motivation of introducing an
additional gauge field is to generate enough deformation of the IR
geometry so as to be dual to an insulating phase in zero temperature
limit. In our previous work \cite{LingEPA}, the coupling term between
the lattice and the electromagnetic field plays a key role in
generating a hard gap in insulating phase. However, at the same
time this coupling term will weaken the influence of the
electromagnetic field to the background such that the IR
deformation due to lattices will becomes weak as well. To solve
this difficulty we intend to consider the two-gauge formalism in
this paper. We will demonstrate that in the presence of the second
gauge field the Q-lattice will strongly deform the IR geometry
even when the coupling term becomes strong such that an
insulating ground state with hard gap can be obtained in zero
temperature limit. In particular, such kind of ground states has
vanishing DC conductivity when the coupling parameter is large
enough. Some novel features of the model such as the non-Drude
behavior will be briefly addressed as well.

\section{Mott-like insulating phase}

We propose the following action as the starting point of our
current model
\begin{eqnarray}
S&=&\frac{1}{2\kappa^2}\int
d^4x\sqrt{-g}[R+\frac{6}{L^2}
-|\nabla\Phi|^2-\frac{m^2}{L^2}|\Phi|^2
-\frac{1}{4}F^2
-\frac{Z(\Phi)}{4}G^2],
\label{action}
\end{eqnarray}
where $L$ is the AdS radius and $F=dA$, $G=dB$ are curvatures of
two $U(1)$ gauge fields $A$ and $B$, respectively. We
will treat $B$ field as the Maxwell field and will concentrate
on its transport properties through this paper. The
coupling term $Z(\Phi)$ is introduced to describe the interaction
between the Q-lattice $\Phi$ and the Maxwell field $B$ in bulk
geometry. Specifically, we will set this coupling as
$Z(\Phi)=(1-\beta|\Phi|^2)^2$ with $\beta$ being positive. It is
worthwhile to notice that  $Z(\Phi)$ is positive definite such
that the stability of system is guaranteed. Previously, a similar
coupling has been introduced in a simple holographic model without
the breaking of translational symmetry in \cite{Mefford:2014gia}.

Consider the following ansatz for the background fields
\fa
ds^2&=&{1\over
z^2}\left[-(1-z)p(z)Udt^2+\frac{dz^2}{(1-z)p(z)U}+V_1dx^2+V_2dy^2\right],\nonumber\\
A&=&\mu(1-z)a dt,\nonumber\\
B&=&\mu(1-z)b dt,\nonumber\\
\Phi&=&e^{i\hat kx}z^{3-\Delta}\phi(z),
\ffa
where $p(z)=1+z+z^2-\mu^2z^3/4$ and
scaling dimension of the scalar field $\Delta=3/2\pm (9/4+m^2)^{1/2}$. For convenience, we have fixed
the radius of AdS $L=1$. All the functions $U,V_1,V_2,a,b$ and
$\phi$ depend on the radial coordinate $z$ only. Through this
paper we set $m^2=-2$ so that
$\Delta=2$. The background solutions to the equations of motions from (\ref{action}) are obtained with full backreaction.
We set the boundary conditions $a(0)=1$ and let $b_0 = b(0)$ be a tunable parameter. The temperature of the dual system is given by
\begin{eqnarray}
\hat T=\frac{(12-\mu^2)U(1)}{16\pi}.
\label{tem}
\end{eqnarray}

Obviously, as $a(z)=1,b(z)=0$ and $\phi(z)=0$ the background
solution is nothing but the standard RN-AdS black hole. For a given coupling constant
$\beta$, each electrically charged black hole solution is
specified by four scaling-invariant parameters, namely, the
dimensionless Hawking temperature $\hat T/\mu$, the lattice amplitude
$\hat\lambda/\mu^{3-\Delta}$ where $\hat\lambda\equiv\phi(0)$, the wave
vector $\hat k/\mu$ and $b_0$. For convenience, we abbreviate these
quantities to $T$, $\lambda$ and $k$, respectively. The periodic
structure with lattice constant $l\equiv 2\pi/k$ is manifest
due to the presence of the complex scalar field $\Phi$, which
plays a key role in visualizing the Mott thought experiment by
holography \cite{LingEPA}.

The electrical conductivity $\sigma(\omega)$ in response to the $B$ field
is obtained with the standard time-dependent
perturbation method in holographic setup. We adopt the following
self-consistent perturbations to the background,
\begin{equation}\label{perturbation}
  \delta A_x = a_x (z)e^{-i\omega t}, \; \delta B_x = b_x(z)e^{-i\omega t}, \;\delta g_{tx}=h_{tx}(z)e^{-i\omega t},\;\delta \Phi=ie^{ikx}z^{3-\Delta}\varphi(z)e^{-i\omega t}.
\end{equation}
Once the background solution is obtained, the corresponding
linearized perturbation equations with variables
$(a_x(z),b_x(z),h_{tx}(z),\varphi(z))$ can be solved numerically.
We remark that both of gauge fields $A$ and $B$ need to be
perturbed for self-consistency. Since we are especially interested
in the transport properties of the Maxwell field $B$, we turn on
the source $b_x(0)$ on the boundary but keep
$a_x(0)=0$. We also use the diffeomorphism and gauge
transformation to ensure that we are extracting the
current-current correlator, which leads to an additional boundary
condition  $\varphi(0)-ik\lambda h_{tx}(0)/\omega =0$ on the
boundary \cite{Donos:2013eha}. On the horizon side, the
ingoing boundary conditions are imposed. When the perturbation
solutions are obtained the electrical conductivity along
$x$-direction can be obtained by
\begin{equation}\label{cond}
\sigma(\omega)=\left.\frac{\partial_z b_x(z)}{i\omega b_x (z)}\right|_{z=0}.
\end{equation}

\begin{figure}
\center{
\includegraphics[scale=0.5]{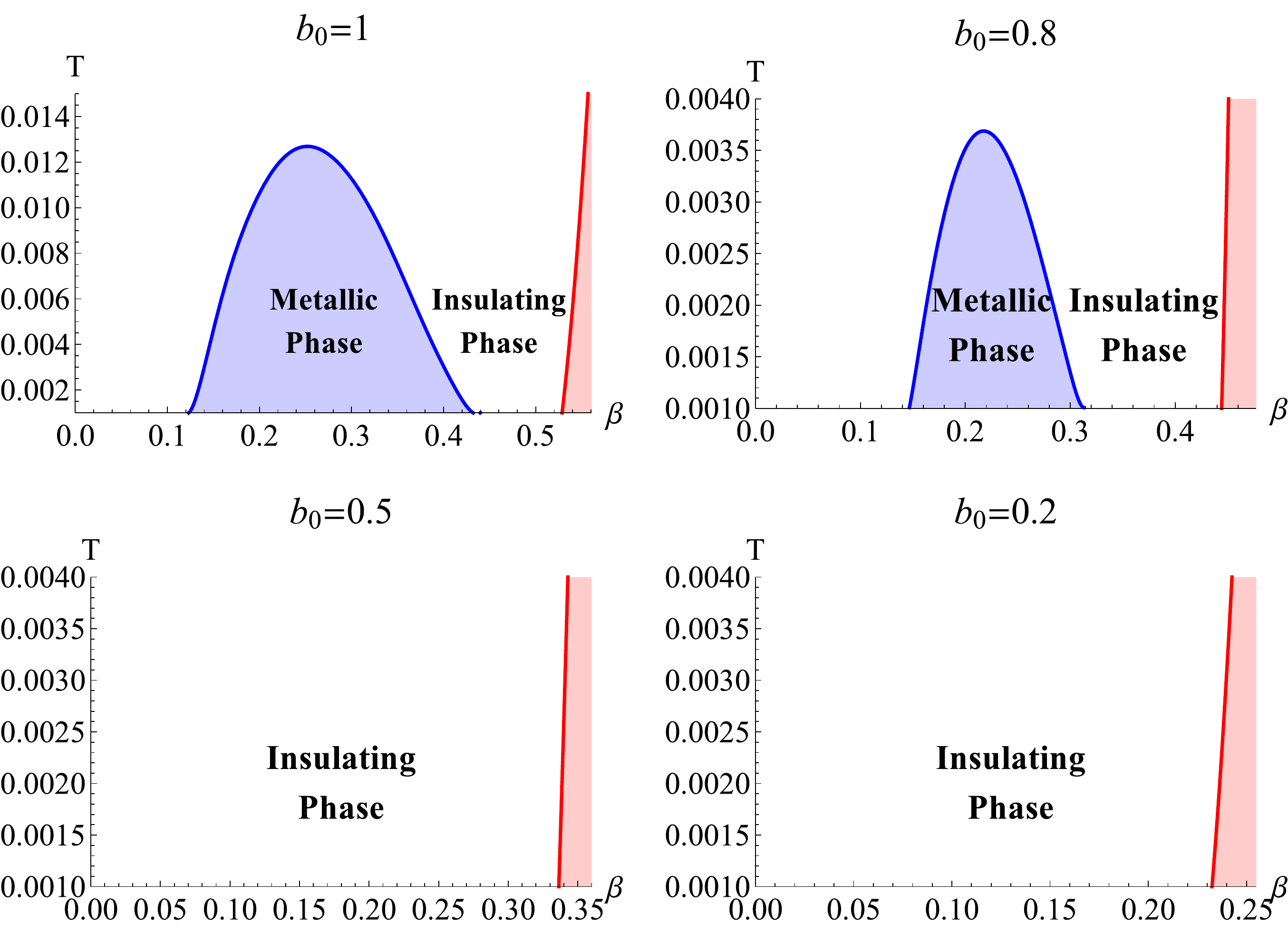}
\caption{\label{phaseall} Phase diagrams with $b_0=1,0.8,0.5,0.2$
respectively from the top left to the right down. It is obviously
seen that as $b_0$ is decreased the metallic phase region
downsizes and eventually vanishes. The insulating and metallic
phases are marked on each plot, the red zone is the no-solution
region, in which the solution to background equations does
not exist and numerically the borderline of this region is
tied to vanishing charge density $\rho\simeq 0$. All the data are
collected above an extremely low temperature $T/\mu=0.001$.}}
\end{figure}

First of all, we intend to demonstrate the phase diagram of the
system when the coupling term $\beta$ is turned on. In the case of
$\beta=0$, it has been known that metal-insulator transition can
occur by adjusting either of parameters $\lambda$ and $k$.
However, a hard gap, which is one of the important
characteristics of most Mott insulator, is absent in the optical
conductivity in this original setup. In our previous work
\cite{LingEPA}, we improve this by introducing the coupling term
$\beta$ such that a hard gap can be manifestly observed as $\beta$
becomes large. However, in zero temperature limit an insulating
phase with hard gap is never observed, which is attributed to the
doped effects from the background. Instead, the system always
exhibits a metallic behavior in zero temperature limit whenever a
hard gap emerges with the increase of $\beta$. Now we intend to
investigate this issue again in the context of two-gauge
formalism, focusing on the transport properties of the dual system
in zero temperature limit. Without loss of generality we fix
$\lambda=2$ and $k=0.03$ which corresponds to an insulating phase
for all values of $b_0$ when $\beta=0$. We plot the phase diagram
over $(T,\beta)$ plane for different values of $b_0$, as
illustrated in Fig. \ref{phaseall}. We remark that in this figure
the insulating phase is determined by $\sigma'_{DC}(T)>0$ whereas
the metallic phase by $\sigma'_{DC}(T)<0$, where the prime denotes
the derivative with respect to $T$, and DC conductivity is
calculated by the horizon data of the background fields
\cite{Donos:2014uba,LingEPA,Ling:2015dma}
\begin{eqnarray}
\sigma_{DC}=\sqrt{\frac{V_2}{V_1}}{\left.\left[Z(\Phi)+\frac{V_1}{2}\left(\frac{Z(\Phi)\mu
b}{k\phi}\right)^2\right]\right|_{z=1}}. \label{DC}
\end{eqnarray}
Equivalently, DC conductivity can be obtained by the
zero frequency limit of AC conductivity from Eq.(\ref{cond}),
which has been confirmed to be identical in our numerics.

A substantial progress shown in this phase diagram is that
there exists an obvious interval labelled by insulating phase
between the region of metallic phase and the region of no-solution
in zero temperature limit, as shown in the first two diagrams with
$b_0=1.0$ and $b_0=0.8$ in Fig. \ref{phaseall}. Moreover, as we
disclosed in \cite{LingEPA}, a hard gap always emerges when
$\beta$ is relatively large. Above observations indicate that an
insulating phase with hard gap could be achieved at zero
temperature in our current system. More surprisingly, when we
decrease the value of $b_0$, the region labelled by metallic phase
shrinks and eventually disappears, as shown in the last two
diagrams with $b_0=0.5$ and $b_0=0.2$ in Fig. \ref{phaseall},
which means that the system will always exhibit insulating
behavior at zero temperature for arbitrary $\beta$. In
\cite{LingEPA} we propose the emergence of metallic phase is
attributed to the doped property of the model, which originates
from the interaction term between Maxwell field and the lattice.
In our current paper it seems that the presence of additional
gauge field $A$ plays a role in suppressing this doped effect.
Phenomenologically, the magnitude of this suppression seems to be
depicted by the ratio of chemical potential of gauge field $A$ to
that of $B$, \emph{i.e.} $1/b_0$, which can be observed from Fig.
\ref{phaseall} since the region of metallic phase shrinks with the
decrease of $b_0$. Nevertheless, the nature of this suppression
asks for further understanding.

\begin{figure}
\center{
\includegraphics[scale=0.55]{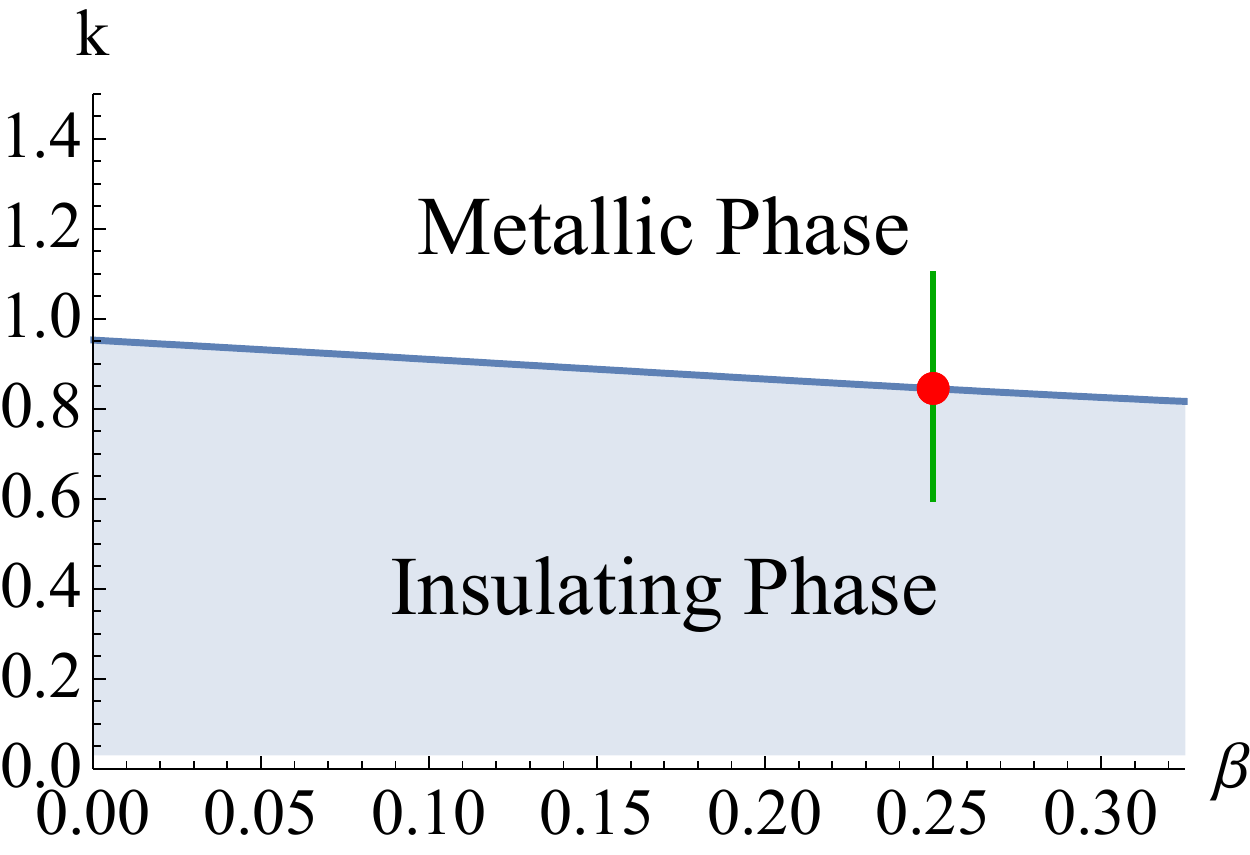}\;\;\;
\includegraphics[scale=0.55]{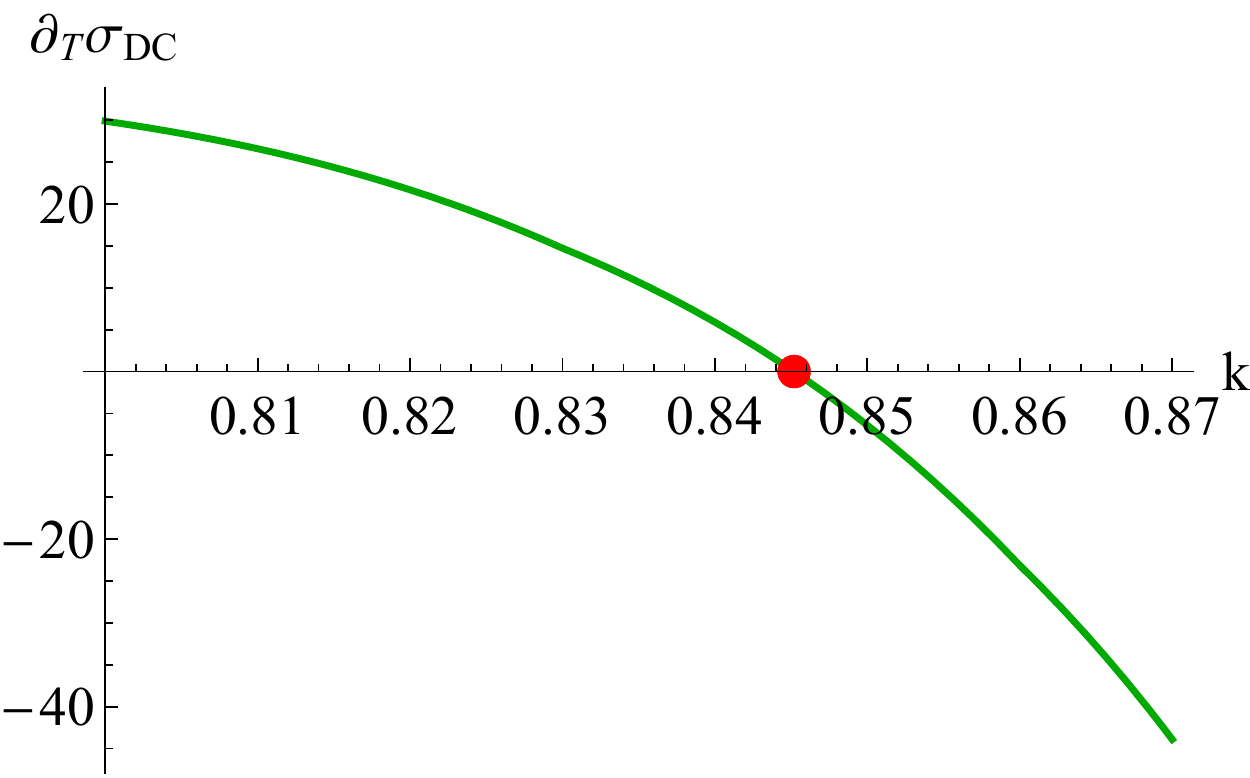}
\caption{\label{kbeta}The left plot is the phase diagram with
($\lambda=2, b_0=0.5, T=0.001$), our numerical data are collected
beyond $k=0.03$. The right plot is the result of
$\partial_{T}\sigma_{DC}\;v.s.\;k$ at $\beta=0.25$ and
other parameters are the same as the left plot. The curve in
right plot corresponds to the segment of the green vertical line
in the left plot, with a critical point labelled by a red dot.}}
\end{figure}

Next we turn to demonstrate the phase diagram over $(k,\beta)$
plane with other parameters fixed. An example is given in Fig.
\ref{kbeta}, manifestly showing that the system undergoes a
transition from a metallic phase to an insulating phase with the
decrease of the wave number $k$, which is just the realization of
the thought experiment proposed by Mott \cite{Mott:1930s}. It is
remarkable that holography can provide us such a powerful
tool to visualize this scenario, as we pointed out in previous
work \cite{LingEPA}. In comparison with the phase diagram
presented in \cite{LingEPA}, one key difference in our current
work is that with appropriate value of $b_0$ (here we set
$b_0=0.5$), the phase diagram over $(k,\beta)$ plane will maintain
its structure down to extremely low temperature with little
change. That is to say, for small values of $k$ the system will
always stay in an insulating phase even in zero temperature limit,
which coincides with our previous results as illustrated in Fig.
\ref{phaseall}. In contrast, the structure of phase diagram
presented in \cite{LingEPA} depends on the temperature evidently,
where we only show an example at normal temperature.

\begin{figure}
\center{
\includegraphics[scale=0.7]{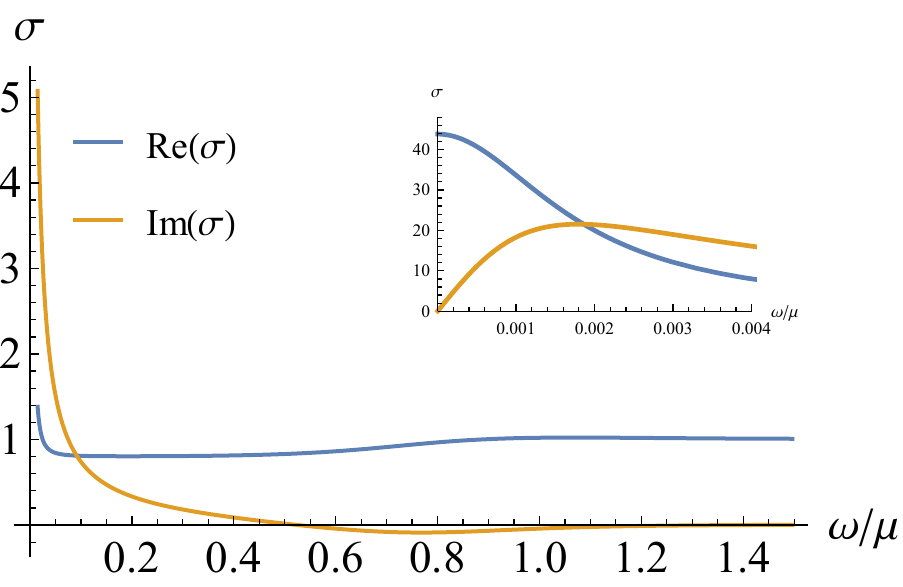}\ \hspace{0.4cm}
\includegraphics[scale=0.8]{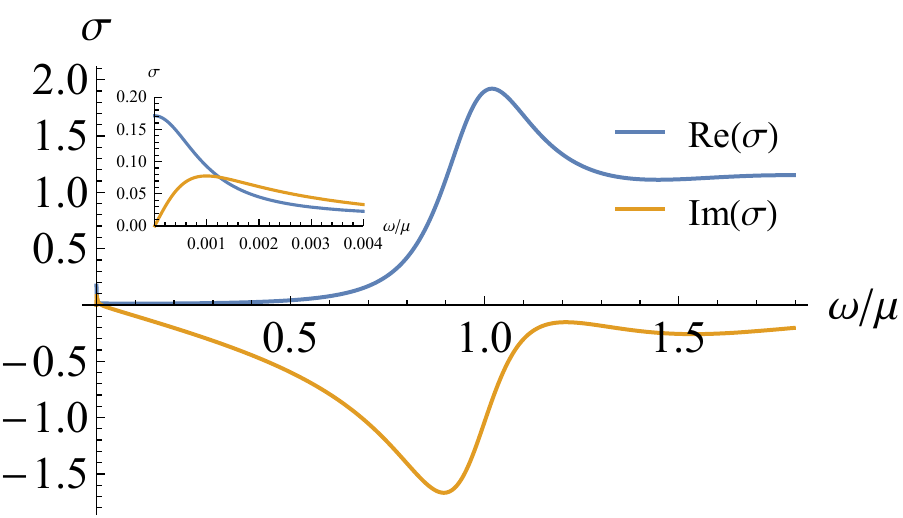}
\caption{\label{sigma}The optical conductivity with different
values of $\beta$ (left plot is for $\beta=0$ and right one for
$\beta=0.34$) at $T=0.005$. The other parameters are fixed as
$\lambda=2$, $k=0.03$, and $b_0=0.5$. The insets in both
plots are the blow-up of the optical conductivity in low frequency
region.}}
\end{figure}

Another key and interesting characteristic of our dual model
is the presence of a hard gap, which is one of the important
properties of most Mott insulators. We demonstrate it in Fig.
\ref{sigma} by computing the frequency dependence of the optical
conductivity. It is noticed that a hard gap emerges and becomes
pronounced when $\beta$ becomes relatively large. We remark
that all above results indicate that the novel insulating
phase emerged at large $\beta$ shares some important
characteristic of a Mott insulator. At phenomenological level,
$\beta$ plays the same role of electron-electron interaction $U$
in Mott-Hubbard model, as has been revealed in \cite{LingEPA}. It
is also observed that the charge density $\rho$ associated with
the Maxwell field $B$ is monotonously decreasing with the
parameter $\beta$, and the borderline of the no-solution region is
characterized by $\rho\simeq 0$. The process of $\beta$
approaching the borderline of no-solution region can be understood
as the process of $U\to\infty$ where all electrons are localized.

\begin{figure}
\center{
\includegraphics[scale=0.7]{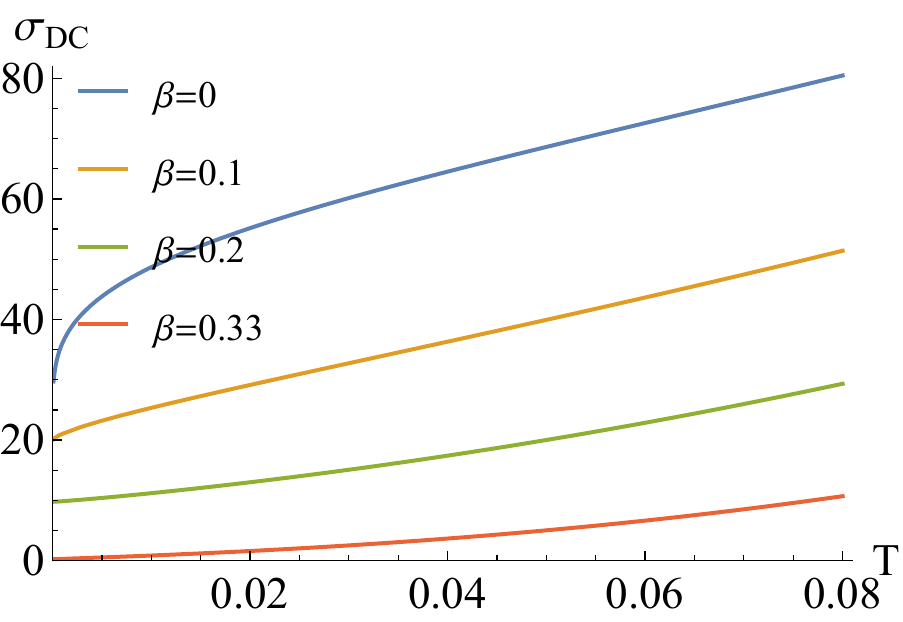} \hspace{1cm}
\includegraphics[scale=0.52]{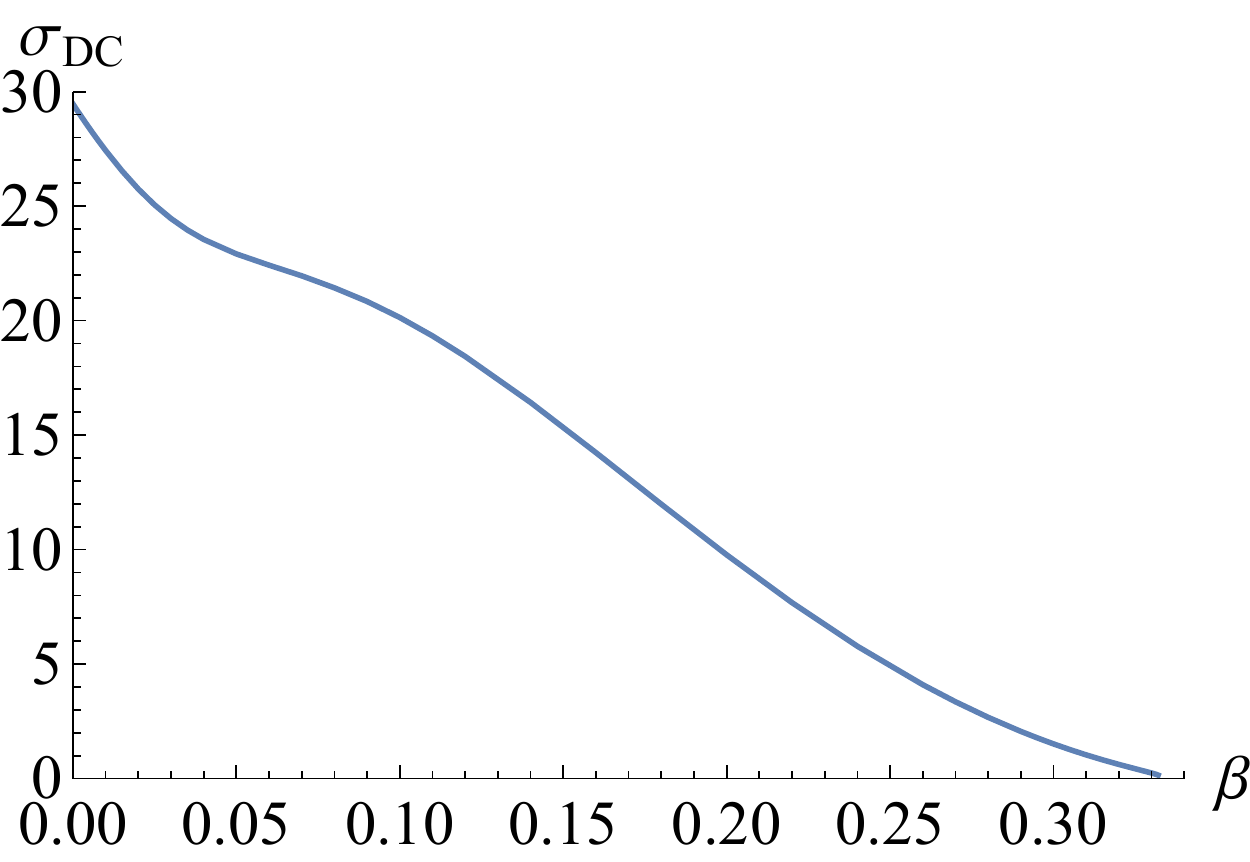}
\caption{\label{DCvsTbeta}Left plot: DC conductivity as the
function of temperature $T$ with different $\beta$. Right plot: DC
conductivity as the function of $\beta$ with fixed temperature
$T=0.0001$. All the other parameters are fixed as $\lambda=2$,
$k=0.03$ and $b_0=0.5$.}}
\end{figure}
Next we turn our interests to the properties of the dual
system in zero temperature limit. First, as we have disclosed in
the phase diagram, when the coupling parameter $\beta$ is large,
the system lies in an insulating phase down to an extremely
low temperature ($T\sim10^{-5}$), which is in contrast to what we
observed in \cite{LingEPA}. Second, we are concerned with the
behavior of the DC conductivity in zero temperature limit. From
the left plot of Fig. \ref{DCvsTbeta}, when the parameter
$\beta$ is small the DC conductivity may approach finite value in
zero temperature limit, and the non-vanishing value decreases with
the increase of $\beta$. When $\beta$ approaches the borderline of
no-solution region, we find that the DC conductivity tends to
vanish as the temperature goes to zero.

\begin{figure}
\center{
\includegraphics[scale=0.8]{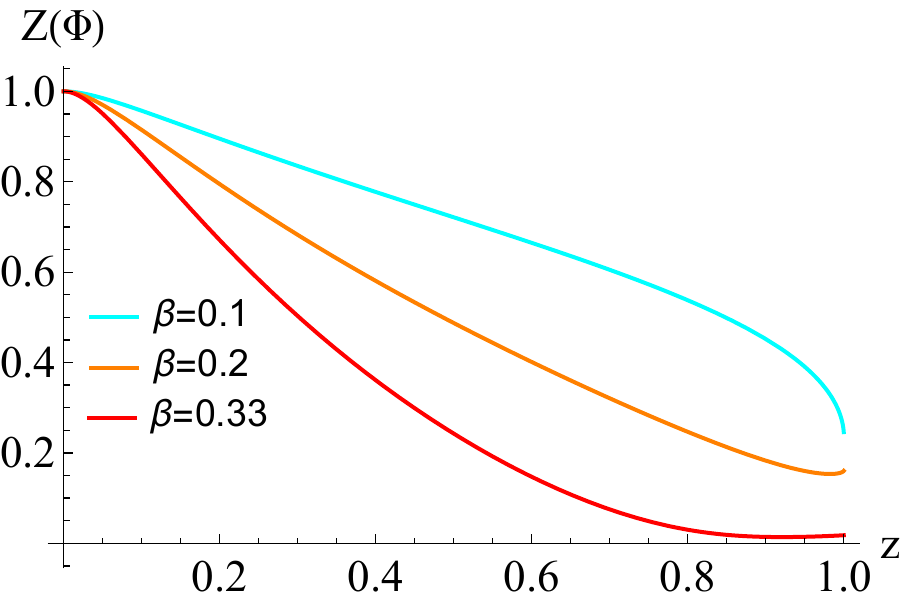}
\caption{\label{Vz}The coupling $Z(\Phi)$ as the function of radial coordinate $z$ with different $\beta$ at low temperature $T=0.0001$.
All the other parameters are fixed as $\lambda=2$, $k=0.03$ and $b_0=0.5$.}}
\end{figure}

In holographic setup it has been understood through the near
horizon analysis that to achieve a vanishing DC conductivity in
zero temperature limit, one essential condition is that the
coupling term $Z(\Phi)$ must be vanishing at horizon as
temperature goes to zero, otherwise this term would contribute
non-zero DC conductivity with both coherent and incoherent parts.
Fig. \ref{Vz} shows the behavior of the coupling term $Z(\Phi)$ as
the function of radial coordinate $z$ with different $\beta$ at
extremely low temperature $T=0.0001$. We find that when $\beta$
approaches the borderline of no-solution region, the near horizon
value of $Z(\Phi)$ tends to zero indeed, implying that DC
conductivity also tends to zero in light of ``membrane paradigm"
of black holes \cite{Iqbal:2008by} (also see Eq. (\ref{DC})).

Therefore, we conclude that we have numerically
obtained a novel insulating phase with hard gap and vanishing DC
conductivity in zero temperature limit, and expect it to point to an
insulating ground state of the system at zero temperature.
Nevertheless, we need to point out with caution that our numerical
analysis above does not guarantee that some new physics might not
be encountered at some ultra low temperature, as also discussed in
the context of Q-lattice setup \cite{Donos:2013eha,Donos:2011ut}.
We expect an analytical treatment to this system could help us to
explore the solidity of this insulating ground state at absolute
zero temperature in future.

\begin{figure}
\center{
\includegraphics[scale=0.5]{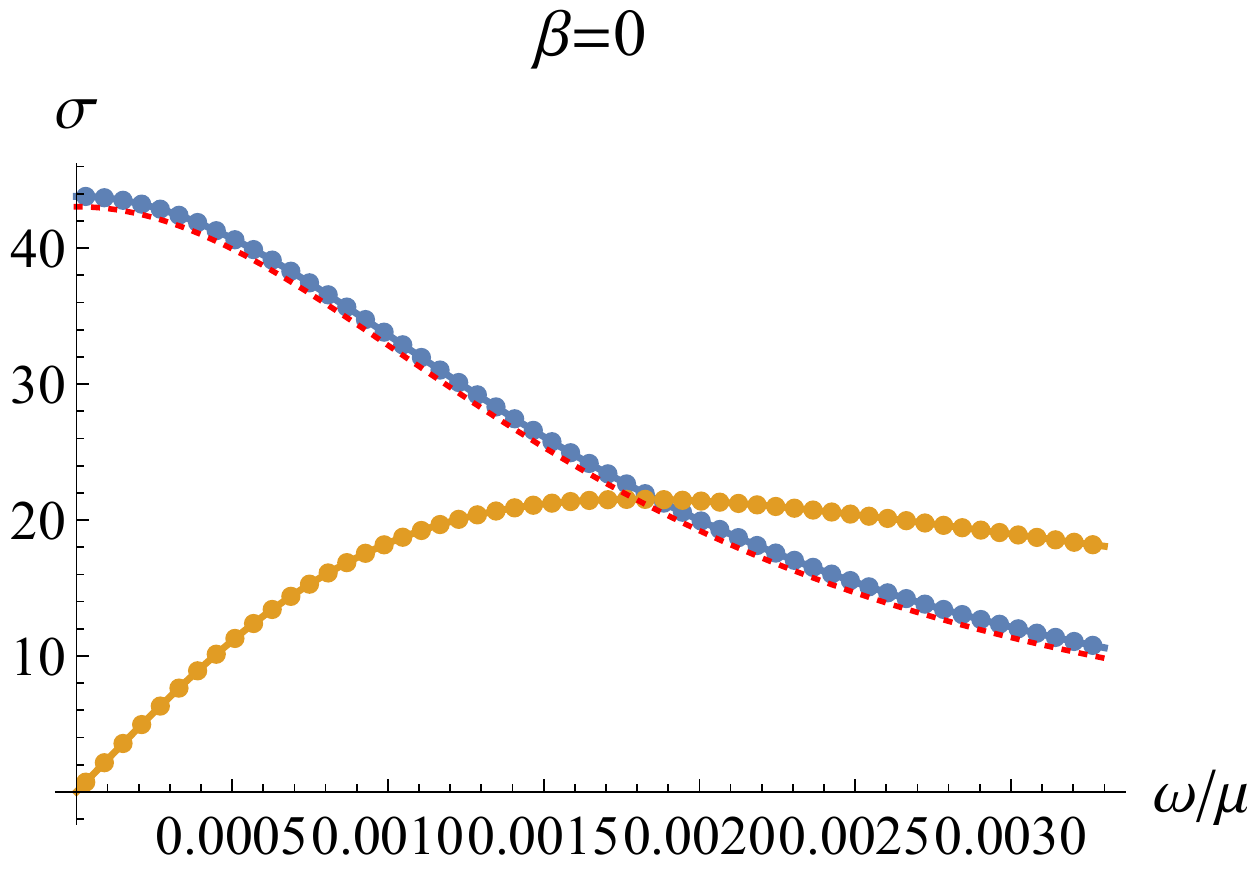}\ \hspace{0.1cm}
\includegraphics[scale=0.5]{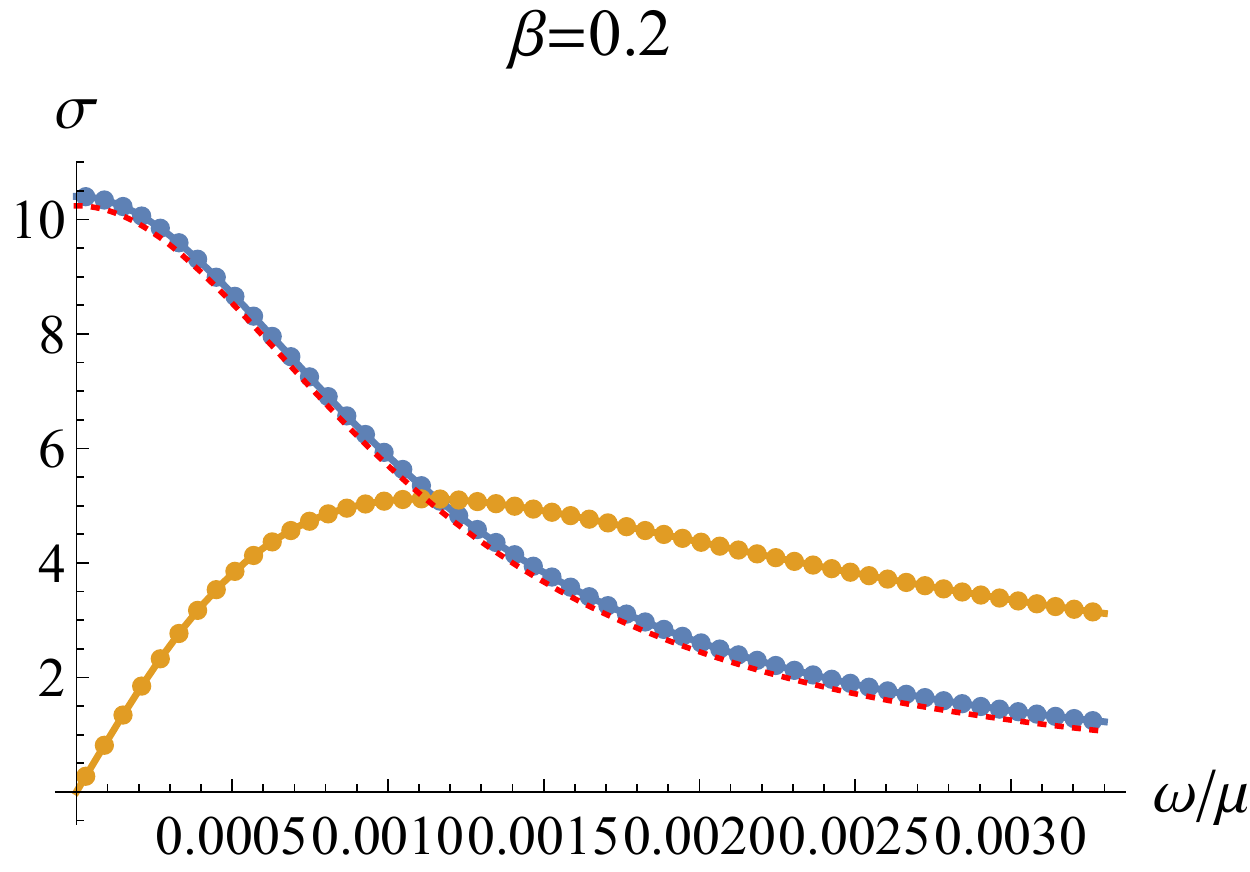}\\ [0.4cm]
\includegraphics[scale=0.5]{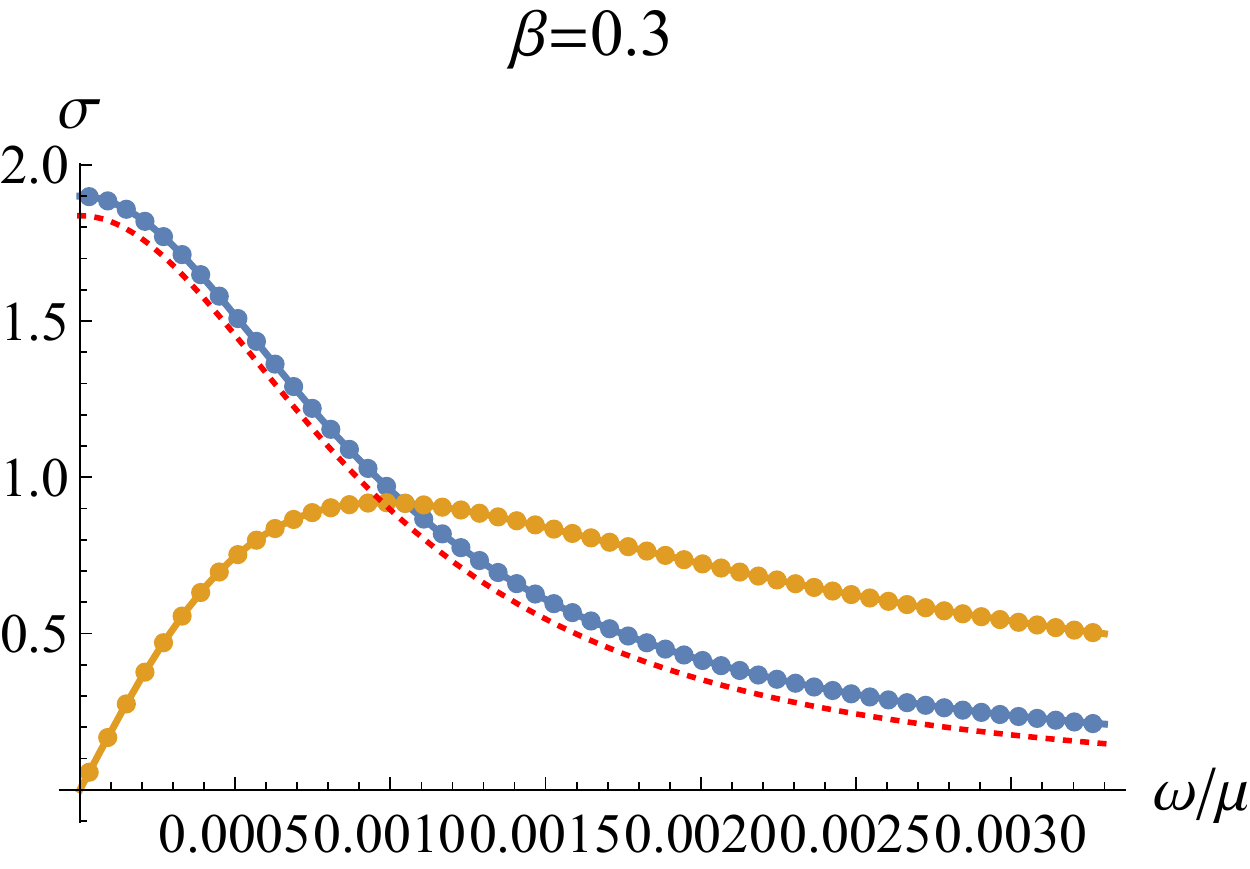}\ \hspace{0.1cm}
\includegraphics[scale=0.5]{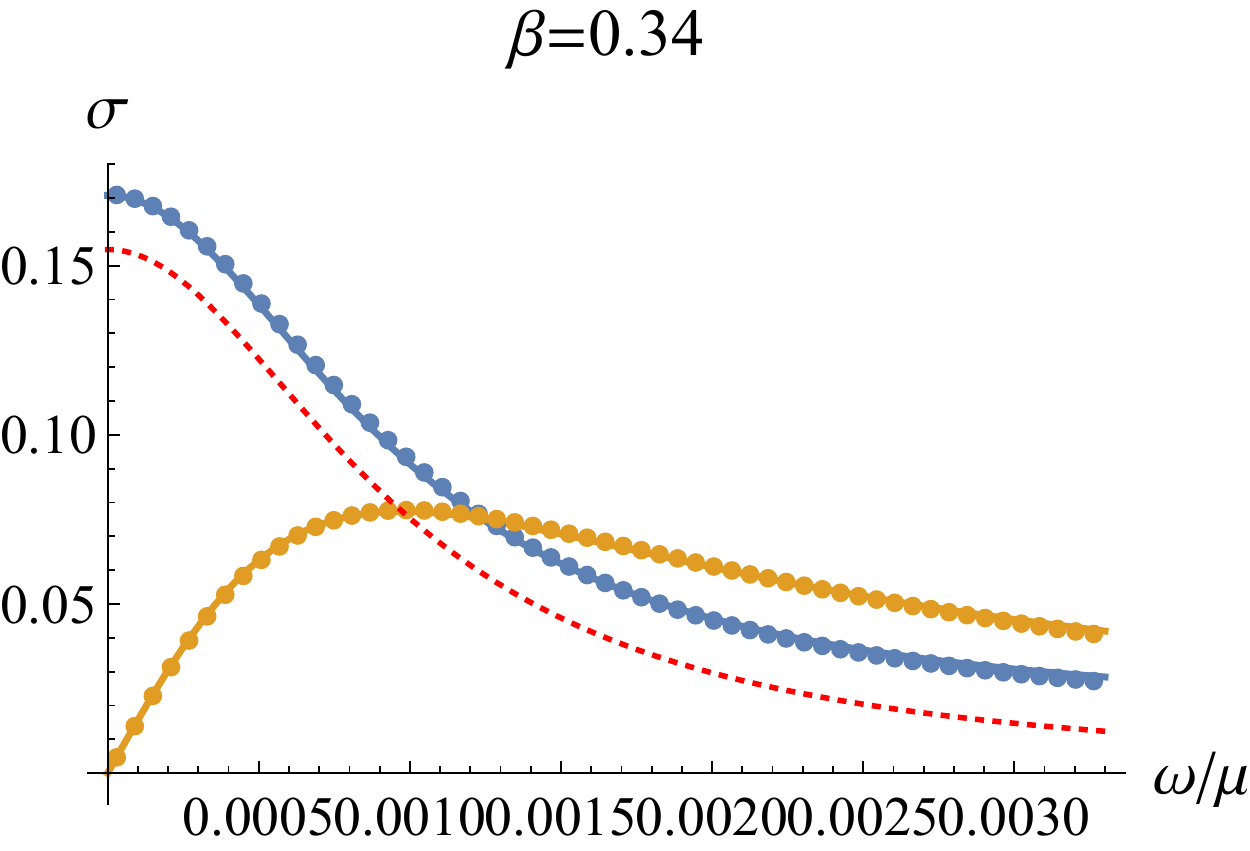}
\caption{\label{sigmalb}The low frequency behavior of the optical
conductivity with $\beta=0, 0.2, 0.3, 0.34$ respectively from the
top left to right down. All the other parameters are fixed as
$\lambda=2$, $k=0.03$ $T=0.005$ and $b_0=0.5$. The red dotted
line in each plot is the real part of $\sigma(\omega)$ when
fitted with the standard Drude formula.}}
\end{figure}

In the end of this section we briefly address the issue on the
zero frequency modes of optical conductivity, which exhibits some
novel characteristics different from ordinary
Mott-insulator\footnote{Recent investigation on the nonvanishing
Drude weight of a Mott insulator can be found, for instance in
\cite{Fujimoto:1998,Nakano:2007}.}. These modes appear simply
because we are observing the system at non-zero
temperature. One example is shown in Fig. \ref{sigmalb}. It can
be found that the conductivity in low frequency region exhibits a
non-Drude behavior in an insulating state. The red dotted
line in Fig. \ref{sigmalb} is a fit with standard Drude formula.
 The
deviation from the Drude law becomes evident with the increase of $\beta$.
In holographic framework, such a discrepancy can be understood as the incoherent contribution which
is decoupled from the momentum relaxation. In particular,
stimulated by recent progress on the incoherence of conductivity
\cite{KimBZA,GeAZA,DavisonTAA,DavisonBEA}, we find our data can be
fitted by a modified Drude formula
\begin{eqnarray}
\sigma(\omega)=\frac{K\tau}{1-i\omega\tau}+\sigma_Q, \label{fit}
\end{eqnarray}
with constant $K$, relaxation time $\tau$ and an additional
constant $\sigma_Q$. $\sigma_Q$ comes from the incoherent
contribution, which is fitted in Table \ref{Tfit}. From both
Fig.\ref{sigmalb} and Table \ref{Tfit}, we notice that for
small $\beta$, $\sigma_Q$ is tiny compared with the
coherent part such that its contribution to the total conductivity can be ignored.
With the increase of $\beta$, both coherent and incoherent
part are suppressed, but finally these two parts are comparable to
each other such that the deviation from Drude law becomes evident.
The suppression of both parts due to the presence of $\beta$ term
also provides us an understanding on the emergence of hard gap and
vanishing DC conductivity in our model.

\begin{widetext}
\begin{table}[ht]
\begin{center}
\begin{tabular}{|c|c|c|c|c|c|c|c|}
         \hline
~$\beta$~ &~$0$~&~$0.1$~&~$0.2$~&~$0.3$~&~$0.34$~
          \\
        \hline
~$\sigma_Q$~ & ~$0.77006$~&~$0.31445$~&~$0.16622$~ & ~$0.06259$~&~$0.01604$~
          \\
        \hline
\end{tabular}
\caption{\label{Tfit}$\sigma_Q$ for different $\beta$ at temperature $T=0.005$. All the other parameters are fixed as $\lambda=2$, $k=0.03$ and $b_0=0.5$.}
\end{center}
\end{table}
\end{widetext}

\section{Discussion}

In this paper we have constructed a Q-lattice model with two
$U(1)$ gauge fields in holographic approach. After introducing a
coupling term to describe the interaction between the lattice and
the Maxwell field, we have demonstrated that the dual system
is a novel insulating phase, which also resembles some important features of a Mott insulator phenomenologically.
Explicitly, the system
undergoes a transition from a metallic phase to insulating phase
as the lattice constant becomes large, which visualizes the
thought experiment proposed by Mott. A hard gap is manifestly
observed as the coupling parameter becomes large. More
importantly, in comparison with our previous work in
\cite{LingEPA}, we have made substantial progress towards the
holographic construction of Mott insulator. We have found the
system stays in an insulating phase with vanishing DC conductivity
in zero temperature limit, implying that the dual system points to
a novel ground state which is insulating rather than a metallic
phase in a doped system as described in \cite{LingEPA}. In optical
conductivity the contribution from incoherent part
becomes evident as the system points to an insulating phase in zero temperature limit.
We find this part could be fitted well with a modified Drude formula which contains an additional constant.

Although we have implemented some important characteristics
of Mott insulator in our novel holographic insulator model,
more investigations are needed to construct a
holographic model dual to a realistic Mott insulator. For
instance, the commensurate nature, a key ingredient of Mott
insulator, has not been addressed. In addition, an analytical
analysis to near horizon geometry is expected to be
implemented, which could provide us more transparent understanding
on the transport properties of the system in zero temperature limit.

It is completely plausible to generalize the construction in
our current simple model to more complicated cases, for instance,
taking the interactions between two gauge fields into account,
or including a charged scalar field \cite{Ling:2014laa} to study the cuprate
phase diagram \cite{Kiritsis:2015hoa,Baggioli:2015dwa}.

\begin{acknowledgments}

We are very grateful to Chao Niu for his collaboration during
the early stage of this work. We also thank the editor and the anonymous referees for
their helpful comments and suggestions. This work is supported by the
Natural Science Foundation of China under Grant Nos.11275208,
11305018 and 11178002. Y.L. also acknowledges the support from
Jiangxi young scientists (JingGang Star) program and 555 talent
project of Jiangxi Province. J. P. Wu is also supported by the
Program for Liaoning Excellent Talents in University (No.
LJQ2014123).

\end{acknowledgments}

\end{document}